# Design Methods for Polymorphic Combinational Logic Circuits based on the Bi_Decomposition Approach

Zhifang Li[1,2], Wenjian Luo[1,2], Lihua Yue[1,2] and Xufa Wang[1,2]

[1] School of Computer Science and Technology, University of Science and Technology of China,
Hefei, 230027, Anhui, China
[2] Anhui Key Laboratory of Software in Computing and Communication, University of Science and Technology of China,
Hefei, 230027, Anhui, China
Email: zhifangl@mail.ustc.edu.cn, {wjluo, llyue, xfwang}@ustc.edu.cn

*Abstract*—Polymorphic circuits are a special kind of digital logic components, which possess multiple build-in functions. In different environments, a polymorphic circuit would perform different functions. Evolutionary Algorithms, Binary Decision Diagrams (BDD) and the multiplex method have been adopted to design polymorphic circuits. However, the evolutionary methods face the scalable problem. The BDD method consumes too much gate resource. The polymorphic circuit built by the multiplex method rarely contains polymorphic gates. In this paper, based on the traditional Bi_Decomposition circuit design approach, two methods, i.e. the Poly_Bi_Decomposition method and the Transformation&Bi_Decomposition method, are proposed for designing polymorphic circuits. The Poly_Bi_Decomposition method can design relatively large and gate-efficient polymorphic circuits with a higher percentage of polymorphic gates. The Transformation&Bi_Decomposition method can use the traditional circuit design approaches and tools, e.g. Bi_Decomposition, to design polymorphic circuits directly. The experimental results show the good performance of the proposed methods.

## I. INTRODUCTION

Polymorphic Electronics are a new research field proposed by Stoica in 2001 [1]. A polymorphic electronic component is sensitive to the environmental signals, and it behaves differently in different environments. For example, a polymorphic NAND/NOR gate controlled by VDD would perform the NAND function when the voltage is 3.3V and perform the NOR function when the Voltage is 1.8V [2]. Current work about Polymorphic Electronics focuses on polymorphic digital logic components, including polymorphic digital logic gates and polymorphic combinational digital logic circuits.

Some polymorphic gates have been designed and fabricated in silicon [1-5], such as the NAND/NOR [2, 4] gate controlled by VDD and the AND/OR [1] gate controlled by temperature. Polymorphic gates are the basic building blocks of polymorphic circuits. Due to the characteristic of multi-functional and sensitiveness to environment signals, polymorphic logic circuits have the potential application in security, verification, multi-functional circuits and smart systems [1].

However, there is no effective method for building large scale polymorphic circuits. Evolutionary Algorithms [3, 6, 7] have been adopted for generating area-efficient polymorphic circuits, but it can not be scaled to large circuits. Up to now, "3×4 multiplier / 7 bit sorting-net" is the largest polymorphic circuit designed by the evolutionary method [8]. Binary Decision Diagrams (BDD) and the multiplex methods have also been used to design polymorphic circuits [9]. However, the BDD method consumes too much gate resource, and the multiplex method hardly utilizes the build-in multi-functional characteristic of polymorphic components.

In this paper, based on the Bi_Decomposition approach for traditional circuit design, the Poly_Bi_Decomposition method for the polymorphic circuit synthesis is proposed. The Poly_Bi_Decomposition method can design gate-efficient and large scale polymorphic circuits with a high percentage of polymorphic gates. In addition, some transformation rules are given for designing polymorphic circuits through existing circuits design method and tools, e.g. Bi_Decomposition [10]. By combining these rules and the Bi_Decomposition method [10], the Transformation&Bi_Decomposition method is proposed.

The rest of this paper is organized as follows. Section II introduces related works briefly. Section III gives a short introduction to the Bi_Decomposition [10]. Section IV explains the proposed methods. Section V demonstrates the experimental results. Section VI gives some discussions. Finally, Section VII concludes the whole paper.

## II. RELATED WORKS

Polymorphic electronic is a novel research field, and several researchers have conducted some pioneer work. In this section, firstly, the works about synthesis of polymorphic circuits through evolutionary methods are summarized. Secondly, the Binary Decision Diagrams (BDD) and polymorphic multiplex methods for designing polymorphic circuits [9] are briefly introduced.

## A. Evolutionary Design of Polymorphic Digital Circuits

Polymorphic gates are the basic building blocks for designing polymorphic digital circuits. The polymorphic gate possesses multiple intrinsically build-in functions. In each mode, the polymorphic gate would perform exactly the same as a traditional logic gate. In [1], Stoica and his colleagues designed the polymorphic gates AND/OR and AND/OR/XOR controlled by VDD, and AND/OR controlled by temperature. Two kinds of NAND/NOR polymorphic gates have been designed and fabricated with the 0.5 and 0.7 CMOS technology in [2] and [4], respectively. The NAND/NOR/NXOR/AND have been reported in [5].

Up to now, there is little theory for guiding the design of polymorphic circuits. In [11] and [12], the definition of complete polymorphic gate sets and the algorithms for judging the completeness of a polymorphic gate set are given, respectively. However, there is no efficient method for guiding the design of large scale polymorphic circuits.

Recently, evolutionary methods are widely adopted for designing polymorphic circuits. In Table I, some polymorphic circuits designed by Evolutionary Algorithms are listed. Currently, the largest polymorphic circuit obtained is "3×4 multiplier / 7 bit sorting-net" [8], which is composed of about 100 polymorphic gates. It can be observed from Table I that it is hard to design large scale polymorphic circuits through evolutionary methods.

TABLE I. POLYMORPHIC CIRCUITS DESIGNED BY EVOLUTIONARY METHODS

| Polymorphic circuits | gates | Generations | Num. of gates | Reference |
|---|---|---|---|---|
| 5-semmetry / 5-median | NAND/NOR, XOR | – | 13 | [13] |
| 2×3multipler / 5-sorting-net | NAND/NOR, AND, OR, XOR | 854,900 | 30 | [8] |
| 3×3multipler / 6-sorting-net | NAND/NOR | 26,972,648 | 52 | [8] |
| 3×3 multiplier / 6-adder | NAND/NOR, OR/XOR | 2,514,043 | 89 | [6] |
| 3×4 multiplier / 7-sorting-net | NAND/NOR, AND | 62,617,151 | 113 | [8] |

## B. Binary Decision Diagrams and Multiplex Methods for Designing Polymorphic Digital Circuits

In [9], the Binary Decision Diagrams (BDD) and polymorphic multiplexes are adopted for designing polymorphic circuits.

As for the PolyBDD method [9], firstly, the original polymorphic function is transmitted to another function. The variable number of the new obtained function is one less than the original function, and its output value is an integer ranged from 0 to 15. Each integer (from 0 to 15) corresponds to a polymorphic component in $\{g_1/g_2 \mid g_1, g_2 \in \{\mathbf{1}, \mathbf{0}, \text{NOT}, \text{WIRE}\}\}$, where $\mathbf{1}$ ($\mathbf{0}$) stands for logic-1 (logic-0) and NOT (WIRE) stands for the NOT (WIRE) logic gate. Secondly, a BDD is generated according to the new obtained function. The internal nodes of the BDD are the variables of the function, and the leaf nodes are integers ranged from 0 to 15. Thirdly, the internal nodes are replaced by multiplexes, and the leaf nodes are replaced by the corresponding component in $\{g_1/g_2 \mid g_1, g_2 \in \{\mathbf{1}, \mathbf{0}, \text{NOT}, \text{WIRE}\}\}$. Finally, a polymorphic circuit implementing the original function is built. The BDD method consumes too much gate resource.

The polymorphic multiplex method [9] is a combination of traditional circuit design method and polymorphic multiplexes. Firstly, traditional circuit design methods, such as ABC [14] and Espresso [15], are adopted to design the single function circuit in each mode. Then those single function circuits are connected to the corresponding input pin of the polymorphic multiplex. The polymorphic multiplex switches one of its inputs to the output according to the environment. The multiplex method could generate gate-efficient results. However, the inherent multifunctional properties of polymorphic gates are not considered, and the circuits designed by this method have no essential different from the traditional multifunctional circuits. It is noted that the polymorphic multiplex is firstly proposed by Sekanina in [13]. Additionally, different kind of polymorphic multiplex based on the complete gate set is given in [11].

## III. INTRODUCTION TO THE BI_DECOMPOSITION APPROACH

The Bi_Decomposition method [10, 16] is an effective approach for designing traditional logic circuits. In [10], Steinbach and Lang give an detailed introduction of the Bi_Decomposition method. In this section, a brief introduction of the Bi_Decomposition circuit design method is given. Figure 1 is an illustration of the principle of the Bi_Decomposition method. The Boolean function $f(A, S, B)$ is the combination of the logic gate $g$ ($g \in \{\text{AND}, \text{OR}, \text{NOT}\}$), the Boolean function $r(A, S)$ and $h(B, S)$. The variable sets A and B would not be empty simultaneously, and $f(A, S, B) = g(r(A, S), h(B, S))$. If neither A nor B is empty, $\{g, r(A, S), h(B, S)\}$ is a strong bi_decomposition. Otherwise, it is a weak bi_decomposition.

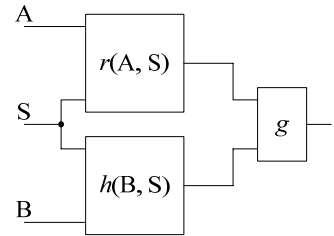

Figure 1. The illustration of the bi_decomposition, where $f(A, S, B) = g(r(A, S), h(B, S))$ [10].

Figure 2 gives an example of the bi_decomposition. The Boolean function $f(x_1, x_2, x_3, x_4)$ in Figure 2(*a*) is decomposed to $r(x_2, x_3, x_4)$ in Figure 2(*b*) and $h(x_1)$ in Figure 2(*c*) through the logic gate AND, i.e. $f(x_1, x_2, x_3, x_4) = \text{AND}(r(x_2, x_3, x_4), h(x_1)\text{f})$. In this instance,

A = {$x_2, x_3, x_4$}, B = {$x_1$} and S = Φ.

The similar decomposition can be conducted to $r$(A, S) and $h$(B, S) until the variable number of the boolean function is not greater than 2. When the decomposition process ends, a circuit implementing the function $f$(A, S, B) is obtained.

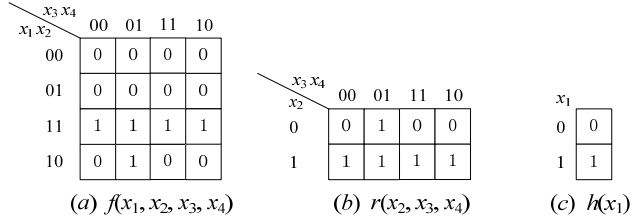

Figure 2. The Boolean function in (*a*) can be decomposed to Boolean functions in (*b*) and (*c*) through the logic gate AND.

## IV. THE PROPOSED METHODS

In this section, based on the bi_decomposition approach [10], two kinds of methods are proposed for designing polymorphic circuits with polymorphic gates as the basic building blocks.

### A. The Poly_Bi_Decomposition method

A polymorphic Boolean function $f$ can be presented as $f_1/f_2$. In mode 1, the function is $f_1$, and in mode 2, the function is $f_2$. For example, the polymorphic Boolean function "4×4multipler / 8bit-sorting-net" performs the function 4×4multipler in mode 1 and function 8 bit sorting-net in mode 2.

Similar to the Bi_Decomposition method in [10], Figure 3 is an illustration of the principle of the Poly_Bi_Decomposition. The polymorphic Boolean function is the combination of a polymorphic gate $g_1/g_2$, polymorphic Boolean functions $r$(A, S) and $h$(B, S), where $g_1, g_2 \in$ {AND, OR, XOR}, $r$(A, S) = $r_1$(A, S)/$r_2$(A, S) and $h$(B, S) = $h_1$(B, S)/$h_2$(B, S). The variable sets A and B would not be empty simultaneously.
$f_1$(A, S, B) = $g_1$( $r_1$(A, S), $h_1$(B, S) ) and
$f_2$(A, S, B) = $g_2$( $r_2$(A, S), $h_2$(B, S) ).

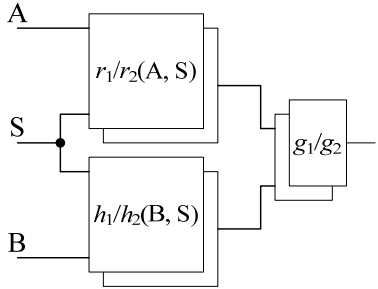

Figure 3. The illustration of the Poly_Bi_Decomposition, where $f_1$(A, S, B) = $g_1$( $r_1$(A, S), $h_1$(B, S) ) and $f_2$(A, S, B) = $g_2$( $r_2$(A, S), $h_2$(B, S) ).

Figure 4 gives an example of the Poly_Bi_Decomposition. The four variable polymorphic Boolean function $f_1/f_2$ is decomposed to a three variable polymorphic Boolean function $r_1/r_2$ and a two variable polymorphic Boolean function $h_1/h_2$. In this example, A={$x_3, x_4$}, B={$x_1$} and S={$x_2$}.

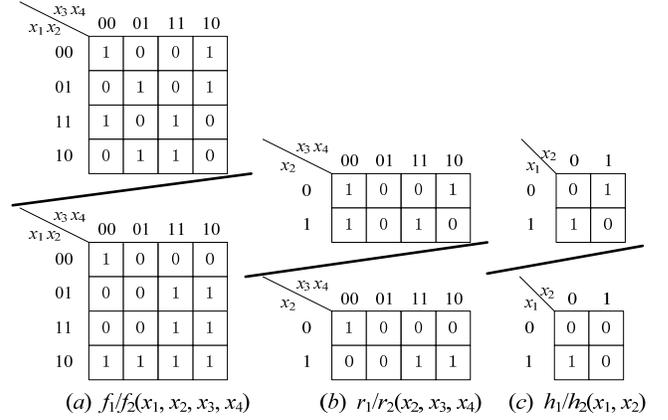

Figure 4. An example of the Poly_Bi_Decomposition, where $f_1$(A, S, B)=XOR($r_1$(A, S), $h_1$(B, S)) and $f_2$(A, S, B)=OR($r_2$(A, S), $h_2$(B, S)).

| Poly_Decomposition ($f_1/f_2$) |
|---|
| *best_decomposition* records the best poly-bi-decomposition. A and B are variable sets. *V* is the input variable set of $f_1/f_2$ |
| in    $f_1/f_2$ is a two modes polymorphic Boolean function |
| out    The polymorphic bi_decomposition of $f_1/f_2$ |
| 1    *best* ← 0, *best_decomposition* ← NULL |
| 2    **for** each $g_1 \in$ {AND, OR, XOR} **do** |
| 3      {$g_2$, A, B} ← find_initial_variable($f_1/f_2$, $g_1$) |
| 4      **if** A ≠ ∅ **then** |
| 5        *Tmp* ← *V* − (A ∪ B) |
| 6        **for** each $x \in$ *Tmp* **do** |
| 7          S ← *V* − (A ∪ B ∪ $x$) |
| 8          **if** there exist $r_1$ and $h_1$ satisfy that $f_1$(A ∪ $x$, S, B) = $g_1$( $r_1$(A ∪ $x$, S), $h_1$(B, S) ) **then** |
| 9            **if** there exist $r_2$ and $h_2$ satisfy that $f_2$(A ∪ $x$, S, B) = $g_2$( $r_2$(A ∪ $x$, S), $h_2$(B, S) ) **then** |
| 10              A ← A ∪ $x$ |
| 11              **goto** step 15 |
| 12          **if** there exist $r_1$ and $h_1$ satisfy that $f_1$(A, S, B ∪ $x$) = $g_1$( $r_1$(A, S), $h_1$(B ∪ $x$, S) ) **then** |
| 13            **if** there exist $r_2$ and $h_2$ satisfy that $f_2$(A, S, B ∪ $x$) = $g_2$( $r_2$(A, S), $h_2$(B ∪ $x$, S) ) **then** |
| 14              B ← B ∪ $x$ |
| 15      **if** | A | > | B | **then** swap(A, B) |
| 16      **if** *best* < | *V* | × min(| A |, | B |) + max(| A |, | B |) **then** |
| 17        *best* ← | *V* | × min(| A |, | B |) + max(| A |, | B |) |
| 18        *best_decomposition* ← {$g_1/g_2$, $r_1/r_2$(A, S), $h_1/h_2$(B, S)} |
| 19    **return** *best_decomposition* |

Figure 5. The algorithm to compute the polymorphic bi_decomposition of the polymorphic Boolean function $f_1/f_2$. The detailed computation process of step 8, step 9, step 12 and step 13 can be found in [10]. The measurement function at step 16 is from [10].

The Poly_Bi_Decomposition of a polymorphic Boolean function $f_1/f_2$ can be computed by the algorithm in Figure 5. In Figure 5, for each $g_1 \in$ {AND, OR, XOR} (step 2), firstly, an initial polymorphic bi-decomposition {$g_1/g_2$, A, B} is obtained by the subroutine "find_initial_variable(…)" at step 3. $g_2$ is different from $g_1$, and it belongs to {AND, OR, XOR}. The size of variable sets A and B are both one. Figure 6 gives the detail of the subroutine

"find_initial_variable(…)". Secondly, a better polymorphic bi-decomposition is generated by the "for" loop started at step 6. Finally, according to the measurement at step 16, a polymorphic bi-decomposition is selected.

| | find_initial_variable($f_1/f_2$, $g_1$) |
|---|---|
| in | $f_1/f_2$ is a two modes polymorphic Boolean function, $g_1 \in$ {AND, OR, XOR}. $V$ is the input variable set of $f_1/f_2$. |
| out | The initial poly-bi-decomposition |
| 1 | **for** each $x_1 \in V$, each $x_2 \in V$ and $x_1 \neq x_2$ **do** |
| 2 | $S \leftarrow V - \{x_1, x_2\}$ |
| 3 | **if** there exist $r$ and $h$ satisfy that $f_1(\{x_1\}, S, \{x_2\}) = g_1( r(\{x_1\}, S), h(\{x_2\}, S) )$ **then** |
| 4 | **if** there exist $g_2$, $r_2$ and $h_2$ satisfy that $f_2(\{x_1\}, S, \{x_2\})=g( r_2(\{x_1\}, S), h_2(\{x_2\}, S) )$ **then** |
| 5 | return $\{g_2, \{x_1\}, \{x_2\}\}$ |
| 6 | **return** {NULL, $\phi$, $\phi$} |

Figure 6. find the initial polymorphic bi-decomposition of $f_1/f_2$.

However, for some polymorphic Boolean functions, the algorithm in Figure 5 cannot give a decomposition. For example, there does not exist a polymorphic bi_decomposition for the polymorphic Boolean function "4 bit parity / 4 bit majority" shown in Figure 7.

Figure 7. The polymorphic Boolean function "4 bit parity / 4 bit majority"

| | Merge&Decomposition($f_1/f_2$) |
|---|---|
| in | $f_1/f_2$ is a polymorphic Boolean function. |
| out | the decomposition of $f_1/f_2$ |
| 1 | The polymorphic Boolean function $f_1/f_2(x_1, \cdots, x_n)$ is transformed to a traditional Boolean function $f'(x_0, x_1, \cdots, x_n)$, where $f'(x_0 = 0, x_1, \cdots, x_n) = f_1(x_1, \cdots, x_n)$ and $f'(x_0 = 1, x_1, \cdots, x_n) = f_2(x_1, \cdots, x_n)$. |
| 2 | The bi_decomposition introduced in [10] is adopted to decompose $f'(x_0, x_1, \cdots, x_n)$. According to the completeness of the bi_decomposition, there always exist variable sets {A′, S′, B′}, a logic gate $g \in$ {AND, OR, XOR} and Boolean functions $\{r', h'\}$ satisfies that $f'(x_0, x_1, \cdots, x_n) = g( r'(A', S'), h'(B', S') )$. |
| 2.1 | If $x_0 \in S'$, let $r_1(A', S'-\{x_0\}) = r'(A', S'-\{x_0\}, x_0 = 0)$, $r_2(A', S'-\{x_0\}) = r'(A', S'-\{x_0\}, x_0 = 1)$, $h_1(B', S'-\{x_0\}) = h'(B', S'-\{x_0\}, x_0 = 0)$ and $h_2(B', S'-\{x_0\}) = h'(B', S'-\{x_0\}, x_0 = 1)$. Therefore, after the decomposition, the new obtained polymorphic Boolean functions are $r_1/r_2$ and $h_1/h_2$.<br>**return** $\{g, r_1/r_2(A, S), h_1/h_2(B, S)\}$ |
| 2.2 | If $x_0 \notin S'$ and $x_0 \in A'$, let $r_1(A'-\{x_0\}, S') = r'(A'-\{x_0\}, S', x_0 = 0)$ and $r_2(A'-\{x_0\}, S') = r'(A'-\{x_0\}, S', x_0 = 1)$. Therefore, after the decomposition, the new obtained functions are the polymorphic Boolean function $r_1/r_2$ and the single mode Boolean function $h'$.<br>**return** $\{g, r_1/r_2(A, S), h'(B, S)\}$ |
| 2.3 | If $x_0 \notin S'$ and $x_0 \in B'$, let $h_1(B'-\{x_0\}, S') = h'(B'-\{x_0\}, S', x_0 = 0)$ and $h_2(B'-\{x_0\}, S') = h'(B'-\{x_0\}, S', x_0 = 1)$. Therefore, after the decomposition, the new obtained functions are the polymorphic Boolean function $h_1/h_2$ and the single mode Boolean function $r'$.<br>**return** $\{g, r'(A, S), h_1/h_2(B, S)\}$ |

Figure 8. The decomposition of a polymorphic Boolean function through a traditional logic gate

When a polymorphic Boolean function $f = f_1/f_2$ cannot be decomposed to two simpler polymorphic Boolean functions through the process in Figure 5, the operations in Figure 8 can be carried out.

Figure 9. The decomposition of the polymorphic Boolean function "4 bit parity / 4 bit majority"

Figure 9 shows the decomposition of the polymorphic Boolean function "4 bit parity / 4 bit majority" in Figure 7. Firstly, the polymorphic Boolean function in Figure 7 is transformed to the single mode Boolean function shown in Figure 9(a). Then, by adopting the bi_decomposition method in [10], the function in Figure 9(a) is decomposed to functions in Figure 9(b) and Figure 9(c) through a OR logic gate. Furthermore, the single mode Boolean functions in Figure 9(b) and Figure 9(c) are transformed to polymorphic Boolean functions in Figure 9(d) and Figure 9(e), respectively.

| | Poly_Design($f_1/f_2$) |
|---|---|
| | $f_1/f_2$ is a polymorphic Boolean function. $V$ is the input variable set of $f_1/f_2$. |
| 1 | **if** $|V| > 2$ **then** |
| 2 | $\{p, r_1/r_2, h_1/h_2\} \leftarrow$ Poly_Decomposition($f_1/f_2$) |
| 3 | **if** $p$ = NULL **then** |
| 4 | $\{g, r, h\} \leftarrow$ Merge&Decomposition($f_1/f_2$) |
| 5 | **if** $r$ is a single mode Boolean function **then** |
| 6 | The bi_decomposition method in [10] is adopted to design a circuit implementing $r$ |
| 7 | **else** |
| 8 | Poly_Design($r$) |
| 9 | **if** $h$ is a single mode Boolean function **then** |
| 10 | The bi_decomposition method in [10] is adopted to design a circuit implementing $h$ |
| 11 | **else** |
| 12 | Poly_Design($h$) |
| 13 | **else** |
| 14 | Poly_Design($r_1/r_2$) |
| 15 | Poly_Design($h_1/h_2$) |

Figure 10. The algorithm to design polymorphic circuits through the polymorphic bi_decomposition method.

For any polymorphic Boolean function, it can be decomposed through the algorithm in Figure 5 or Figure 8. Firstly, it is checked that whether a polymorphic gate can decompose the polymorphic Boolean function to two simpler polymorphic Boolean functions (Figure 5). If the answer is yes, a polymorphic bi_decomposition is obtained. Otherwise, the original polymorphic Boolean function is merged to a single mode Boolean function, and the traditional bi_decomposition approach [10] is adopted to decompose the single mode Boolean function. And the new obtained two single mode Boolean functions are transformed to polymorphic Boolean functions (Figure 8).

Similar to the circuit design process through bi_decomposition in [10], Figure 10 gives the algorithm to design polymorphic circuits through the Poly_Bi_Decomposition.

Figure 11 shows the "4 bit parity / 4 bit majority" polymorphic circuits designed by the algorithm in Figure 10.

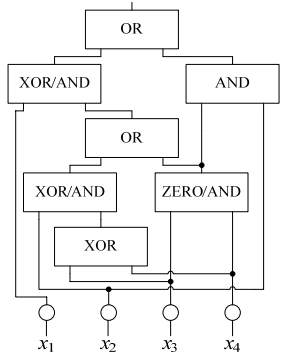

Figure 11. The "4 bit parity / 4 bit majority" polymorphic circuit designed by the polymorphic bi_decomposition method

### B. The Transformation&Bi_Decomposition method

In fact, through a transformation process, the bi_decomposition method in [10] can be used to design polymorphic circuits directly. The steps to design a polymorphic circuit are given below.

(1) The polymorphic Boolean function $f_1/f_2(x_1, \cdots, x_n)$ is transformed to a single mode Boolean function $f'(x_0, x_1, \cdots, x_n)$, where $f'(x_0 = 0, x_1, \cdots, x_n) = f_1(x_1, \cdots, x_n)$ and $f'(x_0 = 1, x_1, \cdots, x_n) = f_2(x_1, \cdots, x_n)$.

(2) The bi_decomposition method in [10] is adopted to design the circuit $Cir$ implementing $f'(x_0, x_1, \cdots, x_n)$.

(3) For every gate $g$ of $Cir$, if $x_0$ or $\overline{x}_0$ is the input of $g$, let $Var_g$ denote the input variable set which influence the output of $g$. Suppose $Cir_g$ is the subcircuit which is composed of $g$ and all the logic gates in $Cir$ that would influence the output of $g$. Let $Cir_g(x_0 = 1)$ denote the function of $Cir_g$ when the value of $x_0$ is logic-1, and $Cir_g(x_0 = 0)$ denote the function of $Cir_g$ when the value of $x_0$ is logic-0.

(3.1) $|Var_g| \leq 3$. Clearly, $Cir_g(x_0 = 0)$ and $Cir_g(x_0 = 1)$ perform as some logic gates, and they are denoted as $g_1$ and $g_2$, respectively. The subcircuit $Cir_g$ is replaced by the polymorphic gate $g_1/g_2$.

(3.2) $|Var_g| > 3$. Without loss of generality, suppose $x_0$ is connected to input pin A of $g$. Suppose the subcircuit connected to the input pin B of $g$ is $Cir_{g\text{-B}}$.

(3.2.1) If $g =$ AND, $g(0, Cir_{g\text{-B}}) = 0$ and $g(1, Cir_{g\text{-B}}) = Cir_{g\text{-B}}$. Therefore, $g$ is replaced by the polymorphic gate ZERO/WIRE.

(3.2.2) If $g =$ OR, $g(0, Cir_{g\text{-B}}) = Cir_{g\text{-B}}$ and $g(1, Cir_{g\text{-B}}) = 1$. Therefore, $g$ is replaced by the polymorphic gate WIRE/ONE.

(3.2.3) If $g =$ XOR, $g(0, Cir_{g\text{-B}}) = Cir_{g\text{-B}}$ and $g(1, Cir_{g-B}) = \overline{Cir_{g-B}}$. Therefore, $g$ is replaced by the polymorphic gate WIRE/NOT.

After the operations in Step 3, the circuit $Cir$ is transformed to a polymorphic circuit implementing the function $f_1/f_2(x_1, \cdots, x_n)$.

The rest part of this section gives the process of constructing the "4 bit parity / 4 bit majority" polymorphic circuit.

Firstly, the polymorphic Boolean function in Figure 7 is transformed to a single mode Boolean function in Figure 9(a). Secondly, the bi_decomposition method in [10] is adopted to design a traditional logic circuit implementing the function in Figure 9(a), and the structure of the circuit is shown in Figure 12. Thirdly, those parts in dashed rectangles are replaced according to the rules given in Step 3. Finally, the polymorphic circuit wanted is obtained, and it is shown in Figure 13.

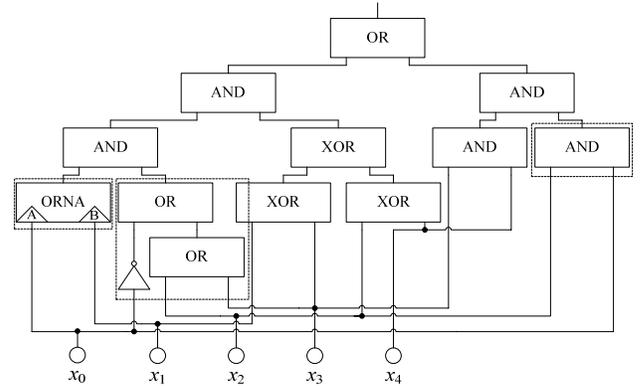

Figure 12. The circuit implementing the Boolean function in Figure 9(a)

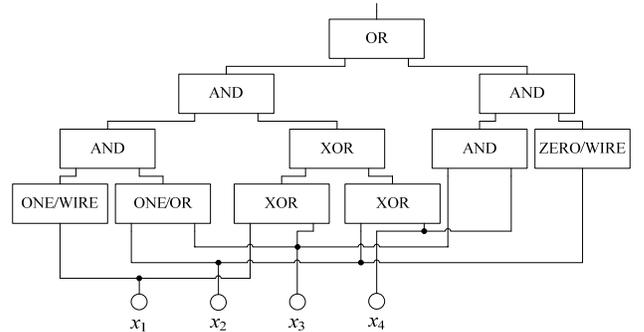

Figure 13. The "4 bit parity / 4 bit majority" polymorphic circuit designed by the Transformation&Bi_Decomposition method

In fact, with the transformation method in Step 1 and Step 3, any traditional circuit design methods (such as ABC [14] and Espresso [15]) can be used to design polymorphic circuits.

For example, a polymorphic circuits implementing $f_1(x_1, \cdots, x_n)/f_2(x_1, \cdots, x_n)$ can be designed by ABC with the following steps. (1) $f_1(x_1, \cdots, x_n)/f_2(x_1, \cdots, x_n)$ is transformed to a single mode Boolean function $f'(x_0, x_1, \cdots, x_n)$, where $f'(0, x_1, \cdots, x_n) = f_1(x_1, \cdots, x_n)$ and $f'(1, x_1, \cdots, x_n) = f_2(x_1, \cdots, x_n)$. (2) The ABC method is used to design the circuit $C$ implementing $f'(x_0, x_1, \cdots, x_n)$. (3) Part of the circuit $C$ is replaced by the corresponding polymorphic gates. At the end, the polymorphic circuit implementing $f_1(x_1, \cdots, x_n)/f_2(x_1, \cdots, x_n)$ is obtained.

## V. EXPERIMENTS

In this section, "parity / majority", "multiplier / sorting-net", and some polymorphic Boolean function constructed by the traditional Boolean function selected from the MCNC [17] library are adopted to test the performance of the proposed methods. The results of the proposed methods are compared with both PolyBDD and polymorphic multiplex methods introduced in [9].

In the design process, both traditional logic gates and polymorphic logic gates are adopted. When a polymorphic Boolean function is decomposed, the traditional logic gates adopted are AND, OR and XOR, and the polymorphic gates are {AND/OR, AND/XOR, OR/AND, OR/XOR, XOR/AND, XOR/OR}.

TABLE II. THE EXPERIMENTAL RESULTS ON "MULTIPLIER / SORTING-NET"

| The number of inputs | 2×3 / 5bit | 3×3 / 6bit | 3×4 / 7bit | 4×4 / 8bit | 5×5 / 10bit | 6×6 / 12bit |
|---|---|---|---|---|---|---|
| PolyBDD [9] | – | – | – | 1028 (–) | – | – |
| Multiplex method based on ABC [9] | 61 (–) | 119 (–) | 198 (–) | 359 (–) | – | – |
| Multiplex method based on Espresso [9] | – | – | – | 2309 (–) | – | – |
| Poly_Bi_Decomposition | 49 (8.6%) | 145 (35.8%) | 248 (27.0%) | 570 (43.8%) | 2507 (36.5%) | 10130 (25.1%) |
| Transformation&Bi_Decomposition | 65 (20.0%) | 170 (22.3%) | 263 (11.0%) | 630 (13.5%) | 2667 (7.7%) | 10329 (5.1%) |

TABLE III. THE EXPERIMENTAL RESULTS ON "PARITY / MAJORITY"

| The number of inputs | 7 | 9 | 11 | 13 | 15 |
|---|---|---|---|---|---|
| Multiplex method based on ABC [9] | 39 (–) | 58 (–) | 79 (–) | 112 (–) | – |
| Poly_Bi_Decomposition | 41 (1) | 59 (2) | 90 (1) | 128 (2) | 186 (1) |
| Transformation&Bi_Decomposition | 64 (5) | 71 (2) | 181 (5) | 144 (2) | 999 (126) |

TABLE IV. THE EXPERIMENTAL RESULTS ON 8 POLYMORPHIC CIRCUITS

|  | majority10 / sao24 (10 → 1) | parity10 / sao24 (10 → 1) | 4×4mul / f 51m (8 → 8) | sorting-net8 / f51m (8 → 8) |
|---|---|---|---|---|
| Poly_Bi_Decomposition | 206 (10.6%) | 54 (40.7%) | 354 (16.1%) | 175 (25.1%) |
| Transformation&Bi_Decomposition | 208 (5.7%) | 119 (11.7%) | 375 (5.3%) | 235 (8.0%) |

|  | ex1010 / sorting_net10 (10 → 10) | 5xp1 / z5xp1 (7 → 10) | 5×5mul / ex1010 (10 → 10) | misex3 / misex3c (14 → 14) |
|---|---|---|---|---|
| Poly_Bi_Decomposition | 2,789 (23.8%) | 98 (60.2%) | 3,587 (40.2%) | 4,571 (48.5%) |
| Transformation&Bi_Decomposition | 3,022 (6.1%) | 152 (13.8%) | 3,716 (9.1%) | 4,682 (7.5%) |

Table II shows the experimental results of "multiplier / sorting-net". In Table II, the number outside the bracket is the number of gates consumed, and the number inside the bracket is the percentage of the polymorphic gates. It can be observed from Table II that the Poly_Bi_Decomposition method consumes less gate resource than the Transformation&Bi_Decomposition method. The difference of the two proposed methods is not large in terms of gate resource. But, the polymorphic gate percentage of the circuits designed by the Poly_Bi_Decomposition method is much higher. The "Multiplex method based on ABC" can build the most gate-efficient polymorphic circuits. However, because the multiplex is used to switch the output of different subcircuits to the multiplex's output, the "Multiplex method based on ABC" do not really utilize the build-in multi-functional property of polymorphic gates.

Table III shows the results of the "parity / majority". In Table III, the number outside the bracket is the number of gates consumed, and the number inside the bracket is number of the polymorphic gates. It can be observed that the "Multiplex method based on ABC" still performs the best in terms of gate resource. The Poly_Bi_Decomposition method is comparable with the "Multiplex method based on ABC".

Table IV shows the experimental results on 8 polymorphic circuits. The circuits, including sao24, 5xp1, z5xp1, ex101, misex3, misex3c and f51m, are taken from the MCNC [17] library. Majority10 is the 10 bit majority Boolean function. Parity10 is the 10 bit parity Boolean function. Sorting-net8 and sorting-net10 are 8 bit and 10 bit sorting-net Boolean function, respectively. It can be observed from Table IV that the Poly_Bi_Decomposition method consumes less gate resource, and the designed circuits have a higher percentage of polymorphic gates. Especially for polymorphic circuits "parity10/sao24" and "5xp1/z5xp1", the gate resource consumed by the Poly_Bi_Decomposition method is much less than those of the Transformation&Bi_Decomposition method, and the percentage of polymorphic gates of the polymorphic circuits designed by Poly_Bi_Decomposition method is much higher.

## VI. DISCUSSIONS

Based on the Bi_Decomposition method in [10], the Poly_Bi_Decomposition method and the Transformation&Bi_Decomposition method are proposed to design polymorphic circuits. The former decomposes the polymorphic Boolean function through a polymorphic gate. The later transforms the polymorphic Boolean function to a single-mode function, and the traditional Bi_Decomposition method is adopted to implement the single mode function. At the end, parts of the circuits are replaced by some polymorphic gates, and the wanted polymorphic circuit is obtained.

Compared with the BDD and multiplex methods in [9], the Bi_Decomposition based methods proposed in this paper can design polymorphic circuits with a higher percentage of polymorphic gates. The BDD method in [9] consumes many multiplexes, and only the lowest level of the circuit consists of polymorphic gates. As for the multiplex method in [9], each single functional subcircuit is designed by traditional logic gates, and polymorphic multiplexes switches one subcircuit's output to the final output. This approach ignores the build-in multi-functional characteristic of polymorphic gates. Oppositely, the Poly_Bi_Decomposition method makes full use of the multi-functional property of polymorphic gates. The polymorphic circuits obtained have a higher percentage of polymorphic gates, and the number of logic gates consumed is relatively reasonable.

The two methods proposed in this paper show their benefits in different aspects. The Poly_Bi_Decomposition method makes full use of the multi-functional property of polymorphic gates, and the circuits designed by the Poly_Bi_Decomposition method have a higher percentage of polymorphic gates. The methodology behind the Transformation&Bi_Decomposition is more universal. In fact, with the transformation method in Section IV-B, any traditional circuit design methods (such as ABC [14]) can be used to design polymorphic circuits. Usually, the percentage of polymorphic gates in circuits designed by the Transformation&Bi_Decomposition method is low, but the traditional circuits design methods with high performance, such as ABC [14], can be used to design polymorphic circuits directly.

## VII. CONCLUSIONS

Based on the Bi_Decomposition method, the Poly_Bi_Decomposition method and the Transformation&Bi_Decomposition method are proposed to design polymorphic circuits. The Poly_Bi_Decomposition method makes full use of the build-in multi-functional property of polymorphic gates, the circuits obtained consumes less gate resource and have a higher percentage of polymorphic gates. Meanwhile, the methodology behind the Transformation&Bi_Decomposition method can be used to design polymorphic circuits through any traditional circuits design method.

ACKNOWLEDGMENT

This work is partly supported by the Fundamental Research Funds for the Central Universities. This paper is based on Zhifang Li's PhD dissertation (in Chinese) in School of Computer Science and Technology at University of Science and Technology of China in June, 2011.